\RequirePackage{lineno}
\documentclass[aps,prd,twocolumn,superscriptaddress,showpacs,amsmath,amssymb]{revtex4}

\usepackage{comment}
\usepackage{multirow}
\usepackage{dcolumn}
\usepackage{bm}
\usepackage{setspace}
\usepackage{graphicx}
\includecomment{printrobustnotes}
\includecomment{printallnotes}
\usepackage{xspace}

\usepackage{dcolumn}
\usepackage{bm}

\newenvironment{cfigure1c}[1][tbp]{\begin{figure*}[#1]\centering}{\end{figure*}}

\newcommand{\note}[1]{\xspace}
\newcommand{\robustnote}[1]{\xspace}
\begin{printrobustnotes}
\renewcommand{\robustnote}[1]{\textbf{\textit{#1}}\xspace}
\end{printrobustnotes}
\begin{printallnotes}
\renewcommand{\note}[1]{\textbf{\textit{#1}}\xspace}
\renewcommand{\robustnote}[1]{\textbf{\textit{#1}}\xspace}
\end{printallnotes}
\newcommand{\ppbar}  {\ensuremath{p\bar{p}}\xspace}
\newcommand{\ttbar}  {\ensuremath{t\bar{t}}\xspace}

\newcommand{\Ht}     {\ensuremath{H_{T}}\xspace}
\newcommand{\et}     {\ensuremath{E_{T}}\xspace}

\newcommand{\met}    {\ensuremath{\text{\raisebox{.3ex}{$\not$}}\et}\xspace}

\newcommand{\mtop}    {\ensuremath{\mathrm{M}_{\text{top}}}\xspace}

\newcommand{\mtnwa}   {\ensuremath{m_{t}^{\text{NWA}}}\xspace}

\newcommand{\mTT}    {\ensuremath{m_{\text{T2}}}\xspace}

\newcommand{\genunit}[2]{\ensuremath{#1~\mathrm{#2}}\xspace}

\newcommand{\gev}[1]    {\genunit{#1}{GeV}}
\newcommand{\pb}[1]     {\genunit{#1}{pb}}

\newcommand{\gevc}[1]   {\ensuremath{#1~\mathrm{GeV}/c}}
\newcommand{\gevcc}[1]  {\ensuremath{#1~\mathrm{GeV}/c^{2}}}

\newcommand{\invfb}[1]  {\ensuremath{#1~\mathrm{fb}^{-1}}}

\newcommand{\gevccnoarg}{\ensuremath{\mathrm{GeV}/c^{2}}\xspace}

\newcommand{\measErr}[2]{\ensuremath{#1 \pm #2}\xspace}
\newcommand{\measAErr}[3]{\ensuremath{#1~^{+#2}_{-#3}}\xspace}

\newcommand{\measStatSyst}[3]{\ensuremath{#1 \pm #2~(\textrm{stat.}) \pm #3~(\textrm{syst.})}\xspace}

\newcommand{\measAStatSyst}[4]{\ensuremath{#1~^{+#2}_{-#3}~(\textrm{stat.}) \pm #4~(\textrm{syst.})}\xspace}

\begin{document}


\title{Measurement of the top quark mass in the dilepton channel using
$\bf{m}_{\text{T2}}$ at CDF}

\affiliation{Institute of Physics, Academia Sinica, Taipei, Taiwan 11529, Republic of China} 
\affiliation{Argonne National Laboratory, Argonne, Illinois 60439} 
\affiliation{University of Athens, 157 71 Athens, Greece} 
\affiliation{Institut de Fisica d'Altes Energies, Universitat Autonoma de Barcelona, E-08193, Bellaterra (Barcelona), Spain} 
\affiliation{Baylor University, Waco, Texas  76798} 
\affiliation{Istituto Nazionale di Fisica Nucleare Bologna, $^{cc}$University of Bologna, I-40127 Bologna, Italy} 
\affiliation{Brandeis University, Waltham, Massachusetts 02254} 
\affiliation{University of California, Davis, Davis, California  95616} 
\affiliation{University of California, Los Angeles, Los Angeles, California  90024} 
\affiliation{University of California, San Diego, La Jolla, California  92093} 
\affiliation{University of California, Santa Barbara, Santa Barbara, California 93106} 
\affiliation{Instituto de Fisica de Cantabria, CSIC-University of Cantabria, 39005 Santander, Spain} 
\affiliation{Carnegie Mellon University, Pittsburgh, PA  15213} 
\affiliation{Enrico Fermi Institute, University of Chicago, Chicago, Illinois 60637}
\affiliation{Comenius University, 842 48 Bratislava, Slovakia; Institute of Experimental Physics, 040 01 Kosice, Slovakia} 
\affiliation{Joint Institute for Nuclear Research, RU-141980 Dubna, Russia} 
\affiliation{Duke University, Durham, North Carolina  27708} 
\affiliation{Fermi National Accelerator Laboratory, Batavia, Illinois 60510} 
\affiliation{University of Florida, Gainesville, Florida  32611} 
\affiliation{Laboratori Nazionali di Frascati, Istituto Nazionale di Fisica Nucleare, I-00044 Frascati, Italy} 
\affiliation{University of Geneva, CH-1211 Geneva 4, Switzerland} 
\affiliation{Glasgow University, Glasgow G12 8QQ, United Kingdom} 
\affiliation{Harvard University, Cambridge, Massachusetts 02138} 
\affiliation{Division of High Energy Physics, Department of Physics, University of Helsinki and Helsinki Institute of Physics, FIN-00014, Helsinki, Finland} 
\affiliation{University of Illinois, Urbana, Illinois 61801} 
\affiliation{The Johns Hopkins University, Baltimore, Maryland 21218} 
\affiliation{Institut f\"{u}r Experimentelle Kernphysik, Karlsruhe Institute of Technology, D-76131 Karlsruhe, Germany} 
\affiliation{Center for High Energy Physics: Kyungpook National University, Daegu 702-701, Korea; Seoul National University, Seoul 151-742, Korea; Sungkyunkwan University, Suwon 440-746, Korea; Korea Institute of Science and Technology Information, Daejeon 305-806, Korea; Chonnam National University, Gwangju 500-757, Korea; Chonbuk National University, Jeonju 561-756, Korea} 
\affiliation{Ernest Orlando Lawrence Berkeley National Laboratory, Berkeley, California 94720} 
\affiliation{University of Liverpool, Liverpool L69 7ZE, United Kingdom} 
\affiliation{University College London, London WC1E 6BT, United Kingdom} 
\affiliation{Centro de Investigaciones Energeticas Medioambientales y Tecnologicas, E-28040 Madrid, Spain} 
\affiliation{Massachusetts Institute of Technology, Cambridge, Massachusetts  02139} 
\affiliation{Institute of Particle Physics: McGill University, Montr\'{e}al, Qu\'{e}bec, Canada H3A~2T8; Simon Fraser University, Burnaby, British Columbia, Canada V5A~1S6; University of Toronto, Toronto, Ontario, Canada M5S~1A7; and TRIUMF, Vancouver, British Columbia, Canada V6T~2A3} 
\affiliation{University of Michigan, Ann Arbor, Michigan 48109} 
\affiliation{Michigan State University, East Lansing, Michigan  48824}
\affiliation{Institution for Theoretical and Experimental Physics, ITEP, Moscow 117259, Russia} 
\affiliation{University of New Mexico, Albuquerque, New Mexico 87131} 
\affiliation{Northwestern University, Evanston, Illinois  60208} 
\affiliation{The Ohio State University, Columbus, Ohio  43210} 
\affiliation{Okayama University, Okayama 700-8530, Japan} 
\affiliation{Osaka City University, Osaka 588, Japan} 
\affiliation{University of Oxford, Oxford OX1 3RH, United Kingdom} 
\affiliation{Istituto Nazionale di Fisica Nucleare, Sezione di Padova-Trento, $^{dd}$University of Padova, I-35131 Padova, Italy} 
\affiliation{LPNHE, Universite Pierre et Marie Curie/IN2P3-CNRS, UMR7585, Paris, F-75252 France} 
\affiliation{University of Pennsylvania, Philadelphia, Pennsylvania 19104}
\affiliation{Istituto Nazionale di Fisica Nucleare Pisa, $^{ee}$University of Pisa, $^{ff}$University of Siena and $^{gg}$Scuola Normale Superiore, I-56127 Pisa, Italy} 
\affiliation{University of Pittsburgh, Pittsburgh, Pennsylvania 15260} 
\affiliation{Purdue University, West Lafayette, Indiana 47907} 
\affiliation{University of Rochester, Rochester, New York 14627} 
\affiliation{The Rockefeller University, New York, New York 10021} 
\affiliation{Istituto Nazionale di Fisica Nucleare, Sezione di Roma 1, $^{hh}$Sapienza Universit\`{a} di Roma, I-00185 Roma, Italy} 

\affiliation{Rutgers University, Piscataway, New Jersey 08855} 
\affiliation{Texas A\&M University, College Station, Texas 77843} 
\affiliation{Istituto Nazionale di Fisica Nucleare Trieste/Udine, I-34100 Trieste, $^{ii}$University of Trieste/Udine, I-33100 Udine, Italy} 
\affiliation{University of Tsukuba, Tsukuba, Ibaraki 305, Japan} 
\affiliation{Tufts University, Medford, Massachusetts 02155} 
\affiliation{Waseda University, Tokyo 169, Japan} 
\affiliation{Wayne State University, Detroit, Michigan  48201} 
\affiliation{University of Wisconsin, Madison, Wisconsin 53706} 
\affiliation{Yale University, New Haven, Connecticut 06520} 
\author{T.~Aaltonen}
\affiliation{Division of High Energy Physics, Department of Physics, University of Helsinki and Helsinki Institute of Physics, FIN-00014, Helsinki, Finland}
\author{J.~Adelman}
\affiliation{Enrico Fermi Institute, University of Chicago, Chicago, Illinois 60637}
\author{B.~\'{A}lvarez~Gonz\'{a}lez$^v$}
\affiliation{Instituto de Fisica de Cantabria, CSIC-University of Cantabria, 39005 Santander, Spain}
\author{S.~Amerio$^{dd}$}
\affiliation{Istituto Nazionale di Fisica Nucleare, Sezione di Padova-Trento, $^{dd}$University of Padova, I-35131 Padova, Italy} 

\author{D.~Amidei}
\affiliation{University of Michigan, Ann Arbor, Michigan 48109}
\author{A.~Anastassov}
\affiliation{Northwestern University, Evanston, Illinois  60208}
\author{A.~Annovi}
\affiliation{Laboratori Nazionali di Frascati, Istituto Nazionale di Fisica Nucleare, I-00044 Frascati, Italy}
\author{J.~Antos}
\affiliation{Comenius University, 842 48 Bratislava, Slovakia; Institute of Experimental Physics, 040 01 Kosice, Slovakia}
\author{G.~Apollinari}
\affiliation{Fermi National Accelerator Laboratory, Batavia, Illinois 60510}
\author{A.~Apresyan}
\affiliation{Purdue University, West Lafayette, Indiana 47907}
% ...

\collaboration{CDF Collaboration\footnote{With visitors from $^a$University of Massachusetts Amherst, Amherst, Massachusetts 01003,
$^b$Universiteit Antwerpen, B-2610 Antwerp, Belgium, 
$^c$University of Bristol, Bristol BS8 1TL, United Kingdom,
$^d$Chinese Academy of Sciences, Beijing 100864, China, 
$^e$Istituto Nazionale di Fisica Nucleare, Sezione di Cagliari, 09042 Monserrato (Cagliari), Italy,
$^f$University of California Irvine, Irvine, CA  92697, 
$^g$University of California Santa Cruz, Santa Cruz, CA  95064, 
$^h$Cornell University, Ithaca, NY  14853, 
$^i$University of Cyprus, Nicosia CY-1678, Cyprus, 
$^j$University College Dublin, Dublin 4, Ireland,
$^k$University of Edinburgh, Edinburgh EH9 3JZ, United Kingdom, 
$^l$University of Fukui, Fukui City, Fukui Prefecture, Japan 910-0017
$^m$Kinki University, Higashi-Osaka City, Japan 577-8502
$^n$Universidad Iberoamericana, Mexico D.F., Mexico,
$^o$University of Iowa, Iowa City, IA  52242,
$^p$Kansas State University, Manhattan, KS 66506
$^q$Queen Mary, University of London, London, E1 4NS, England,
$^r$University of Manchester, Manchester M13 9PL, England,
$^s$Muons, Inc., Batavia, IL 60510, 
$^t$Nagasaki Institute of Applied Science, Nagasaki, Japan, 
$^u$University of Notre Dame, Notre Dame, IN 46556,
$^v$University de Oviedo, E-33007 Oviedo, Spain, 
$^w$Texas Tech University, Lubbock, TX  79609, 
$^x$IFIC(CSIC-Universitat de Valencia), 56071 Valencia, Spain,
$^y$Universidad Tecnica Federico Santa Maria, 110v Valparaiso, Chile,
$^z$University of Virginia, Charlottesville, VA  22906
$^{aa}$Bergische Universit\"at Wuppertal, 42097 Wuppertal, Germany,
$^{bb}$Yarmouk University, Irbid 211-63, Jordan
$^{jj}$On leave from J.~Stefan Institute, Ljubljana, Slovenia, 
}}
\noaffiliation


\date{\today}
\begin{abstract}
We present measurements of the top quark mass using 
$\mTT$, a variable related to the transverse mass in
events with two missing particles.  We use the template method 
applied to \ttbar dilepton events produced in $\ppbar$ 
collisions at Fermilab's Tevatron Collider and collected by the CDF detector.
From a data sample corresponding to an integrated luminosity of
\invfb{3.4}, we select 236 \ttbar candidate events.  Using the $\mTT$
distribution, we measure the top quark mass to be $\mtop  =
\gevcc{\measAStatSyst{168.0}{4.8}{4.0}{2.9}}$.  By combining 
$\mTT$ with the reconstructed top quark mass distributions based on a neutrino
weighting method, we measure $\mtop  =
\gevcc{\measStatSyst{169.3}{2.7}{3.2}}$.  This is the first
application of the $\mTT$ variable in a mass measurement at a hadron collider. 
\end{abstract}

\maketitle
\section{Introduction}
Models in numerous, well-motivated theoretical frameworks make
predictions for
new phenomena at hadron colliders such as the Tevatron and the 
Large Hadron Collider~(LHC)~\cite{TeV4LHC,LHCPhys}. 
Within each framework, one can 
construct a number of qualitatively different models consistent with data. 
Thus, when discoveries are made at a hadron collider, we face the inverse 
problem of how one maps back to the underlying theory 
responsible for the new phenomena~\cite{TeV4LHC,theory_disc}. A potentially powerful observable 
to discriminate among models and to extract the mass of new particles, 
when the new phenomenon produces a pair of new particles with large missing energy 
signatures, is the $\mTT$ variable~\cite{mT2First,mT2nd}. The $\mTT$ variable is based on transverse mass in events with two missing particles.

The top quark is the heaviest known elementary particle with a mass
approximately 40 times larger than the mass of its isospin partner, the bottom quark~($b$). 
The large top quark mass~(\mtop) produces significant
contributions to electroweak
radiative corrections. Therefore, top quark mass measurements 
are important tests of the standard
model and provide constraints on the Higgs boson mass. In
the dilepton channel, \ttbar pair production is followed by the decay
of each top quark to a $W$ boson and a $b$ quark where both $W$ bosons
then decay to charged leptons~($e$ or $\mu$) and neutrinos. Events 
in this channel thus contain two leptons, two $b$ quark jets, and two
undetected neutrinos. 
The measurement of \mtop using complementary techniques
tests and improves our understanding of this important parameter
in the standard model~\cite{PDGtop}.  

In this letter, we present the
first measurement of the mass of the top quark using the $\mTT$
distribution with \ttbar events in the dilepton channel~\cite{mT2LHC}. 
We use this channel because it has decay products similar to
possible new phenomena where undetected particles are created. We
compare this method with two others that 
were previously used: the reconstructed top quark mass using the
neutrino weighting algorithm
(\mtnwa)~\cite{PhysRevD.60.052001,dilmassprd} and the scalar sum of
transverse energies of jets, leptons, and missing transverse 
energy~($\met$)~\cite{coordinate} in the event ($\Ht$)~\cite{Abe:1995ar}. 
We also measure the top quark mass using pairs of observables
(($\mTT$, \mtnwa) and (\mtnwa, $\Ht$)) simultaneously.

\section{The $\bf{m}_{\text{T2}}$ variable}
Many models contain
heavy, strongly-interacting particles with the same conserved 
charge or parity that result in weakly-interacting, stable particles in the
final state. 
A hadron collider would pair-produce these colored particles, which
then decay
into standard model particles along with a pair of undetectable
weakly interacting particles, so that the generic experimental signature 
is large missing transverse momentum accompanied by multiple energetic jets and
leptons~\cite{coordinate}.  In this final state, we can define $\mTT$ as
\begin{linenomath*}
\begin{align}
{\mTT(m_{invis})} \qquad = \qquad \qquad \qquad\qquad \qquad \qquad \qquad \qquad  \notag\\
\min_{\mathbf{p}_{T}^{(1)}, \mathbf{p}_{T}^{(2)}} \left[ {\max [m_T(m_{invis}; \mathbf{p}_{T}^{(1)}),
m_{T}(m_{invis}; \mathbf{p}_{T}^{(2)})]} \right],
\label{eq_min}
\end{align}
\end{linenomath*}
where $m_{T}$, the transverse mass of each parent particle, is defined as 
\begin{linenomath*}
\begin{align}
m_T(m_{invis}; \mathbf{p}_T^{invis}) \qquad = \qquad \qquad \qquad \qquad \qquad \notag\\
 \sqrt{m_{vis}^2 + m_{invis}^2 + 2
(E_{T}^{vis} E_{T}^{invis} -
\mathbf{p}_{T}^{vis} \cdot \mathbf{p}_{T}^{invis})}.\label{eq_mT} 
\end{align}
\end{linenomath*}
Here $invis$ and $vis$ represent the individual undetected~(invisible) and
detected~(visible)
particles, respectively,
$\mathbf{p}_T^{(1)}$ and $\mathbf{p}_T^{(2)}$ are transverse momenta
of two invisible particles and $m_{invis}$ is the mass of the
invisible particle. The minimization is performed with the
constraint
$\mathbf{p}_{T}^{(1)}+\mathbf{p}_{T}^{(2)}=\mathbf{p}_{T}^{missing}$,
where the magnitude of $\mathbf{p}_{T}^{missing}$ is constrained to
the missing transverse momentum.

The quantity $\mTT$ represents a lower bound on the mass of the parent
particle.
Using the $\mTT$ distribution, we can extract the mass of this parent particle~\cite{mT2Kong} 
in a similar way to the precise measurement of the $W$ boson
mass~\cite{WMass} where an event contains one charged 
lepton ($e$ or $\mu$) and a neutrino, with the latter
not being detected.

\section{Experiment and Data}
We use a sample of \ttbar candidates in the dilepton channel, corresponding
to \invfb{3.4} of proton-antiproton collisions at $\sqrt{s}=1.96$~TeV, 
collected using the CDF II detector~\cite{r_cdfDetector}.  
This is a general-purpose detector designed
to study \ppbar collisions at the Fermilab Tevatron.
A charged-particle tracking system, consisting of a silicon
microstrip tracker and a drift chamber, is immersed
in a 1.4~T magnetic field. Electromagnetic and hadronic
calorimeters surround the tracking system and measure
particle energies. Drift chambers and scintillators, located
outside the calorimeters, detect muon candidates. 

We select events consistent with the \ttbar dilepton decay topology.
We require two oppositely charged lepton candidates with 
$p_{T} > \gevc{20}$ with one isolated~\cite{isolated} lepton candidate in the central region~($|\eta|<1$)
of the detector,
and another isolated or non-isolated lepton candidate in the central region, or
isolated electron candidate in the forward region~($1.0<|\eta|<2.0$).  
We also require $\met$ exceeding
\gev{25}, and at least two jets with $E_{T} > \gev{15}$ and
$|\eta|<2.5$~\cite{coordinate}. To further reject backgrounds, we 
request $\Ht > \gev{200}$. We also require the variables of interest
to be consistent with the top quark hypothesis by demanding 
$\gevcc{20}<\mTT<\gevcc{300}$ and $\gevcc{100}<\mtnwa<\gevcc{350}$.
The criteria select 236 \ttbar 
candidate events. 

The primary sources of background production are Drell-Yan, diboson,
and QCD multijet events. 
We estimate the rate of the Drell-Yan events with a
calculation based on simulated events using the {\sc
alpgen}~\cite{r_alpgen} v2.10 Monte Carlo~(MC) 
generator and the rate of diboson events
with a {\sc pythia}~\cite{pythia6.4} v6.216 calculation.
For the Drell-Yan $Z$+jets process, we normalize the MC sample 
by matching the number of $Z$ events predicted and observed in the  
$Z$ mass region between 76 \gevccnoarg and 106 \gevccnoarg. 
We use data to estimate the rate of background events from QCD
multijet  
production where an event has one real lepton and one of the jets
misidentified as another lepton~(fake).  
In measuring the top quark mass, we divide the \ttbar 
candidate sample into events with and without secondary vertex $b$
tags~\cite{r_run2topxsec}, which have very different purity.  We only attempt to $b$ tags the two highest $E_{T}$ jets.  
Table~\ref{table_DILbackground} summarizes the composition 
of background events and the expected numbers of \ttbar and background events.
We estimate the \ttbar signal event rates using {\sc
pythia} v6.216 with {\sc cteq5l}~\cite{refcteq5l} parton distribution functions at leading order with a full detector simulation~\cite{Gerchtein:2003ba}.

\begin{table}
\begin{center}
\caption[Expected numbers of background and signal events]{
Expected and observed numbers of signal and background events 
assuming \ttbar production cross section $\sigma_{\ttbar}$ = \pb{6.7
\pm 0.8} and \mtop =
\gevcc{175}. Uncertainties quoted capture the uncertainties 
on the theoretical cross section, the statistics of data in the $Z$ mass window, the jet energy scale, the luminosity, the fake rates, and the statistics of the MC samples. 
}
\begin{ruledtabular}
\begin{tabular}{ccc}
\label{table_DILbackground}
            &non-tagged         &tagged\\
\hline
Diboson     &15.2~$\pm$~2.3     &0.6 $\pm$ 0.1\\
Drell-Yan   &31.1~$\pm$~3.5     &1.7 $\pm$ 0.2\\
QCD multijets        &31.2~$\pm$~8.7     &4.5 $\pm$ 1.3\\
\hline
Total Background&77.5 $\pm$ 9.8 &6.8 $\pm$ 1.3 \\
\hline
\ttbar with $\sigma_{\ttbar}$ = \pb{6.7} &68.7 $\pm$ 6.8 &88.4 $\pm$ 8.2\\
\hline
Total (Predicted) &146.2 $\pm$ 11.9 &95.2 $\pm$ 8.3\\
\hline
Observed (\invfb{3.4}) &149 &87 \\
\end{tabular}
\end{ruledtabular}
\end{center}
\end{table}

\begin{figure}
\includegraphics[width=0.44\textwidth]{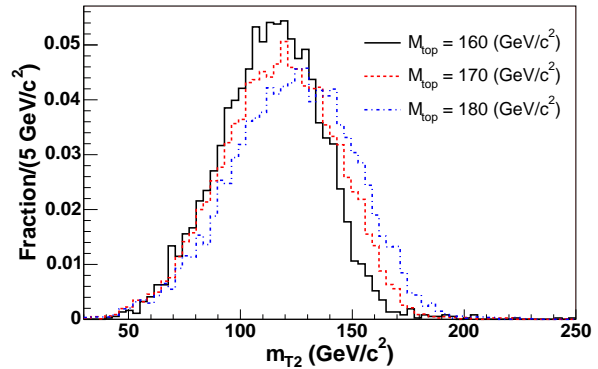}
\caption[$\mTT$ templates in MC]{The $\mTT$ distributions from \ttbar 
dilepton Monte Carlo events that pass the selection criteria for 
three input values of the top quark mass. Each distribution is
normalized to have unit area.}
\label{ref:templatesmt2}
\end{figure}

To calculate $\mTT$ of a $\ttbar$ dilepton event~\cite{mT2LHC}, we
first identify all possible configurations corresponding to different
assignments of jets to $b$ quarks and combinations of
quarks and leptons. The two most energetic jets in an event are
considered to have originated from the $b$ quarks. 
For each configuration,
we calculate the transverse mass of each top quark~($t \rightarrow
bl\nu$) using Eq.~\ref{eq_mT}:
\begin{linenomath*}
\begin{equation}
m_{T} =\sqrt{
m^{2}_{bl}+m^{2}_{\nu}+2(E^{bl}_{T}E^{\nu}_{T}-\mathbf{p}^{bl}_{T}
\cdot \mathbf{p}^{\nu}_{T})},
\end{equation}
\end{linenomath*}
where $m_{bl}$ and $\mathbf{p}^{bl}_T$ denote the invariant mass and
transverse momentum of the bottom-quark jets and charged lepton ($bl$)
system, $m_{\nu}$ and 
$\mathbf{p}^{\nu}_T$ are the mass and transverse momentum of the neutrino, and
$E^{bl}_T$ and $E^{\nu}_T$ are the transverse energies of the $bl$
system and neutrino:
\begin{linenomath*}
\begin{equation}
E^{bl}_T =\sqrt{|\mathbf{p}_{T}^{bl}|^2 +m^2_{bl}}~~\mbox{and}~~E^{\nu}_T =
\sqrt{|\mathbf{p}^{\nu}_T|^2+m^2_{\nu}}.
\end{equation}
\end{linenomath*}
We then calculate \mTT using Eq.~\ref{eq_min} with the assumption 
$m_{\nu}=0$, and for all possible parton assignments. 
We select the smallest value for each event. Figure~\ref{ref:templatesmt2} shows simulated $\mTT$
distributions for various top quark masses for the combined non
$b$-tagged and $b$-tagged sample, which demonstrates that $\mTT$ is
sensitive to \mtop, and thus can be used to measure it.  

\section{Mass fit}
We estimate the probability density functions~(PDFs) of signals and
background using the kernel density
estimation~(KDE)~\cite{KDEHEP,TMTPRD2008} that constructs the PDF without 
any assumption of a functional form. For the mass measurement with two 
observables, we use the two dimensional KDE that accounts for the 
correlation between the two observables.
First, at discrete values of \mtop from \gevcc{130} to \gevcc{220} with increments of \gevcc{0.5} in the region immediately above and below \gevcc{175} to \gevcc{5} near the extreme mass values, we estimate the PDFs for the observables
from 76 $\ttbar$ MC samples. Each sample consists of 0.6M to 4.8M
generated events, 
with 1M events corresponding to a luminosity of 
\invfb{150}, assuming a $\ttbar$ cross section of \pb{6.7}~\cite{ttcros}. 
We smooth and
interpolate the MC distributions 
to find PDFs for arbitrary values of \mtop using the local polynomial smoothing
method~\cite{LPSbook}.  We fit the distributions of the observables in the data 
to the signal and background PDFs in an unbinned extended maximum
likelihood fit~\cite{eml} where
we minimize the negative logarithm of the likelihood using {\sc minuit}~\cite{minuit}. 
The likelihood is built for the $b$-tagged and non $b$-tagged categories separately and then combined by multiplying the two categories. 
We find the statistical uncertainty on \mtop by searching for the points where the
negative logarithm of the likelihood minimized with respect to all
other parameters deviates by 0.5 units from the minimum.
Ref.~\cite{TMTPRD2008} provides detailed information about this
technique.

We test the mass fit procedures using 3,000 pseudo-experiments 
for each of 14 different top quark masses ranging from \gevcc{159} to \gevcc{185} with almost \gevcc{2} step size. 
In each experiment, we select the numbers of background events from a
Poisson distribution with a mean equal to the expected numbers of background events 
in the sample and the numbers of signal events from a Poisson distribution
with a mean equal to the expected numbers of signal events assuming a $\ttbar$ pair 
production cross section of 6.7~pb. 
The distributions of the average mass residual~(deviation from the
input top mass) and the width of the pull~(the ratio of the residual to the
uncertainty reported by {\sc minuit}) for simulated
experiments show that the measured top quark mass is on average
\gevcc{0.26 \pm 0.10}
lower than the true top quark mass and has no dependence on \mtop 
in the $\mTT$ measurements.  We correct the measurement for this bias. 
No such bias is observed with the combined ($\mTT$, \mtnwa)
measurement. In all cases, the fit on average
correctly estimates the statistical uncertainties, based on the pull width
distribution being consistent with unity.  
For \mtop=\gevcc{175}, we expect the statistical uncertainties on \mtop to be 
\gevcc{4.0} with $\mTT$, 
\gevcc{3.4} with \mtnwa, 
\gevcc{5.4} with $\Ht$, 
\gevcc{2.9} with ($\mTT$, \mtnwa) combined, and 
\gevcc{3.2} with (\mtnwa, $\Ht$) combined.

\section{Systematic Uncertainties}
\begin{table*}[ht]
\begin{center}
\caption[Summary of systematics]{Estimated
statistical~(\mtop=\gevcc{175}), 
systematic, and total uncertainties in \gevccnoarg.}
\begin{ruledtabular}
\begin{tabular}{ll|ccc|cc}
\label{systtablesummary}
 &  &$\mTT$ & \mtnwa & $\Ht$ & (\mtnwa, $\mTT$) & (\mtnwa, $\Ht$) \\
\hline
\hline
Statistical & 				     &4.0  & 3.4 &5.4 & 2.9 & 3.2  \\
\hline
Systematic & Jet energy scale~(light quarks) & 2.6 &3.5 &3.7  &3.0   &3.4 \\
	& Generator                          & 0.3 &1.0 &2.6  &0.5   &1.3\\
	& Parton distribution functions      & 0.5 &0.6 &1.8  &0.5   &0.8\\
	& $b$ Jet energy scale               & 0.2 &0.3 &0.2  &0.2   &0.3\\
	& Background shape                   & 0.4 &0.3 &0.7  &0.1   &0.3\\
	& Gluon fusion fraction              & 0.3 &0.1 &0.3  &$<$0.1&0.1\\
	& Initial and final state radiation  & 0.6 &0.2 &0.6  &0.3   &0.2\\
	& MC statistics                      & 0.3 &0.3 &0.5  &0.3   &0.3\\
	& Lepton energy                      & 0.6 &0.2 &0.7  &0.3   &0.2\\
	& Multiple hadron interaction        & 0.2 &0.3 &0.3  &0.3   &0.3\\
	& Color reconnection                 & 0.7 &0.6 &2.5  &0.6   &0.6\\
\cline{2-7}                                           
	& Total systematic uncertainty       & 2.9 &3.8 &5.7  &3.2   &3.8\\
\hline
Total &					     & 5.0 &5.1 &7.8  &4.3   &5.0 \\
\end{tabular}
\end{ruledtabular}
\end{center}
\end{table*}

We examine a variety of systematic effects that could affect the measurement
by comparing MC simulated experiments in which we vary relevant parameters 
within their systematic uncertainties. The dominant source of
systematic uncertainty is the light quark jet energy
scale~(JES)~\cite{r_NIMJES}. We vary JES parameters within their
uncertainties in both signal and background MC generated events and
interpret the
shifts as uncertainties. 
The $b$-jet energy scale systematic uncertainty arising from our
modeling of
$b$ fragmentation, $b$ hadron branching fractions, 
and calorimeter response captures the additional uncertainty not taken
into account in the light quark jet energy scale.  
The uncertainty arising from the choice of MC generator is estimated
by comparing MC simulated experiments generated with 
{\sc pythia} and {\sc herwig}~\cite{r_herwig}. 
We estimate the 
systematic uncertainty due to modeling of initial-state gluon
radiation~(ISR) and final-state gluon radiation~(FSR) by extrapolating
uncertainties in the $p_{T}$ of Drell-Yan events to the \ttbar mass
region~\cite{Abulencia:2005aj}.  
We estimate the systematic uncertainty due to parton distribution functions
by varying the independent eigenvectors of the
{\sc cteq6m}~\cite{refcteq6m} parton distribution functions, varying $\Lambda_{QCD}$, and comparing {\sc cteq5l}~\cite{refcteq5l} with {\sc mrst72}~\cite{refmrst} parton distribution
functions. 
In estimating the systematic uncertainty associated with uncertainties in the top quark
production mechanism, we vary the fraction of top quarks produced by gluon-gluon 
annihilation from 6\% to 20\%, corresponding to the one standard deviation upper bound on
the gluon fusion fraction~\cite{theory_csection}.
We estimate systematic uncertainties due to 
the lepton energy and momentum scales by propagating shifts
in electron energy and muon momentum scales within their uncertainties. 
Background shape systematic uncertainties account for the variation 
of the background composition. In addition, we change 
the shape of the Drell-Yan background sample according to the difference in the
missing energy distribution observed in data and simulation, and 
the shape of the QCD multijet model. 
We estimate the multiple hadron
interaction systematic uncertainties to account for the fact that the
average
number of interactions in our MC samples are not equal to the number
observed in the data. We extract the mass dependence on the number of
interactions in MC pseudo-experiments by dividing our MC samples into
subsamples with different number of interactions. We then
multiply the slope of the result by the difference in the number of
interactions between MC events and data and treat that as a systematic
uncertainty. 

It has been suggested that color reconnection~(CR) effects could cause
a bias in the top quark mass measurement and interpretations at the level of
\gevcc{0.5}~\cite{Wicke:2008iz}. 
We estimate uncertainties arising from CR effects using the {\sc
pythia} 6.4 MC generator, which
includes CR effects and other new features in modeling 
the underlying event, initial and final state radiation, and parton showering.
We generate two MC samples, one using tune A~\cite{Field:2005sa},
which is very similar to the tune for CDF nominal MC generations, the
other using A$_{CR}$~\cite{Wicke:2008iz}, which includes CR into the tune A.
We take the difference 
in the extracted mass between these two MC samples as a systematic uncertainty. 
We measure the difference to be 0.6~GeV for ($\mTT$, \mtnwa) combined, and
0.7~GeV for $\mTT$ alone.
As a cross check, we generate two other MC samples, one using tune
S$_{0}$~\cite{Wicke:2008iz} and the other using NOCR~\cite{Wicke:2008iz}, 
which include all of the new features with and without CR.  
We find a similar mass difference between the two samples. 

Table~\ref{systtablesummary} summarizes the sources and estimates of 
systematic uncertainties. The total systematic uncertainties, adding
them in quadrature, are 
\gevcc{2.9} with $\mTT$, 
\gevcc{3.8} with \mtnwa, 
\gevcc{5.7} with $\Ht$, 
\gevcc{3.2} with ($\mTT$, \mtnwa) combined, and 
\gevcc{3.8} with (\mtnwa, $\Ht$) combined.
The \mTT method has a jet energy scale uncertainty significantly smaller 
than \mtnwa, resulting in the smallest total systematic uncertainty.
Including both statistical and systematic uncertainties, we conclude that 
$\mTT$ is one of the best observables for the \mtop measurement  
, comparable to the measurement using \mtnwa. Using both $\mTT$ and 
\mtnwa, we expect to achieve a 10\% improvement in overall uncertainty
over using $\mTT$ alone.

\section{Results}
We apply a likelihood fit to the data using observables discussed in
this article.
Figure~\ref{ref:loglikelihood} shows the one-dimensional log-likelihoods 
for $\mTT$ and ($\mTT$, \mtnwa) combined. Figure~\ref{ref:fit} shows the 
distributions of the observables used for the \mtop measurements overlaid with 
density estimates using \ttbar signal events with \mtop = \gevcc{169} and 
the full background model.
The fit results are summarized in Table~\ref{datafit}. The extracted masses are 
consistent with each other and the statistical uncertainties 
are consistent with predictions from MC pseudo-experiments.

\begin{cfigure1c}
\begin{tabular}{cc}
\includegraphics[width=0.44\textwidth]{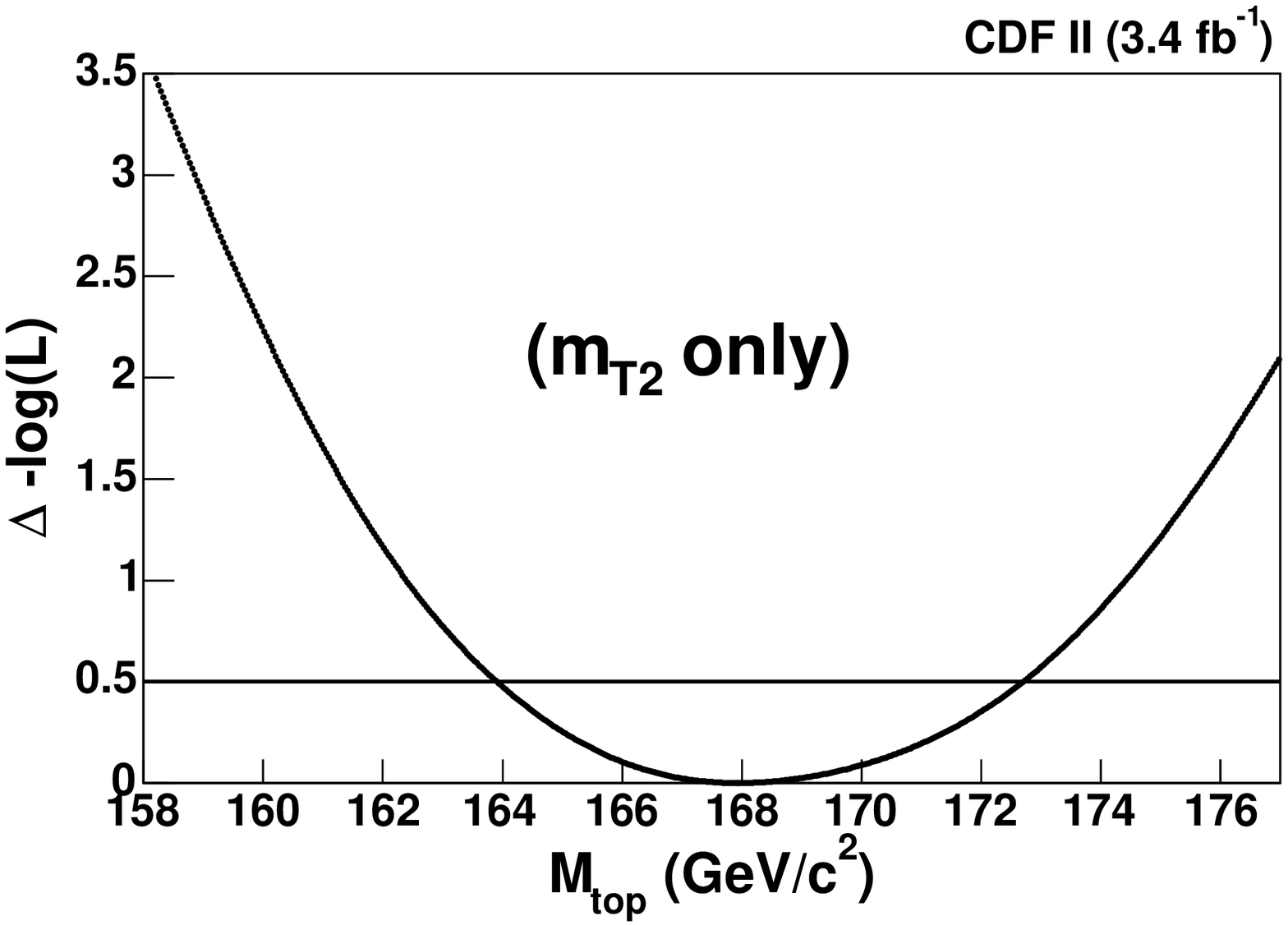}&
\includegraphics[width=0.44\textwidth]{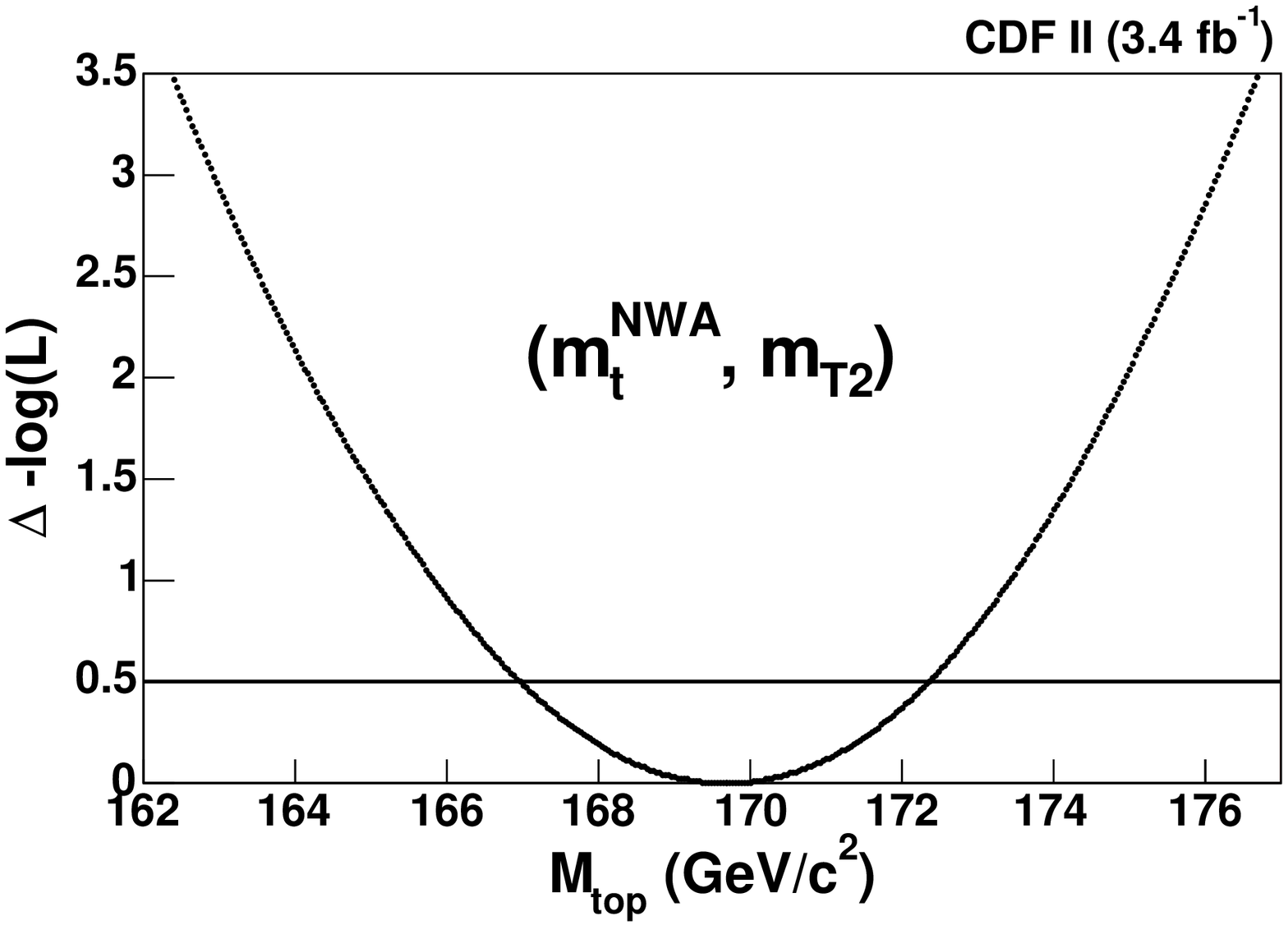}\\
\end{tabular}
\caption[Negative log likelihood]{Negative log-likelihood
distributions. 
}
\label{ref:loglikelihood}
\end{cfigure1c}

\begin{cfigure1c}
\begin{tabular}{ccc}
\includegraphics[width=0.3\textwidth]{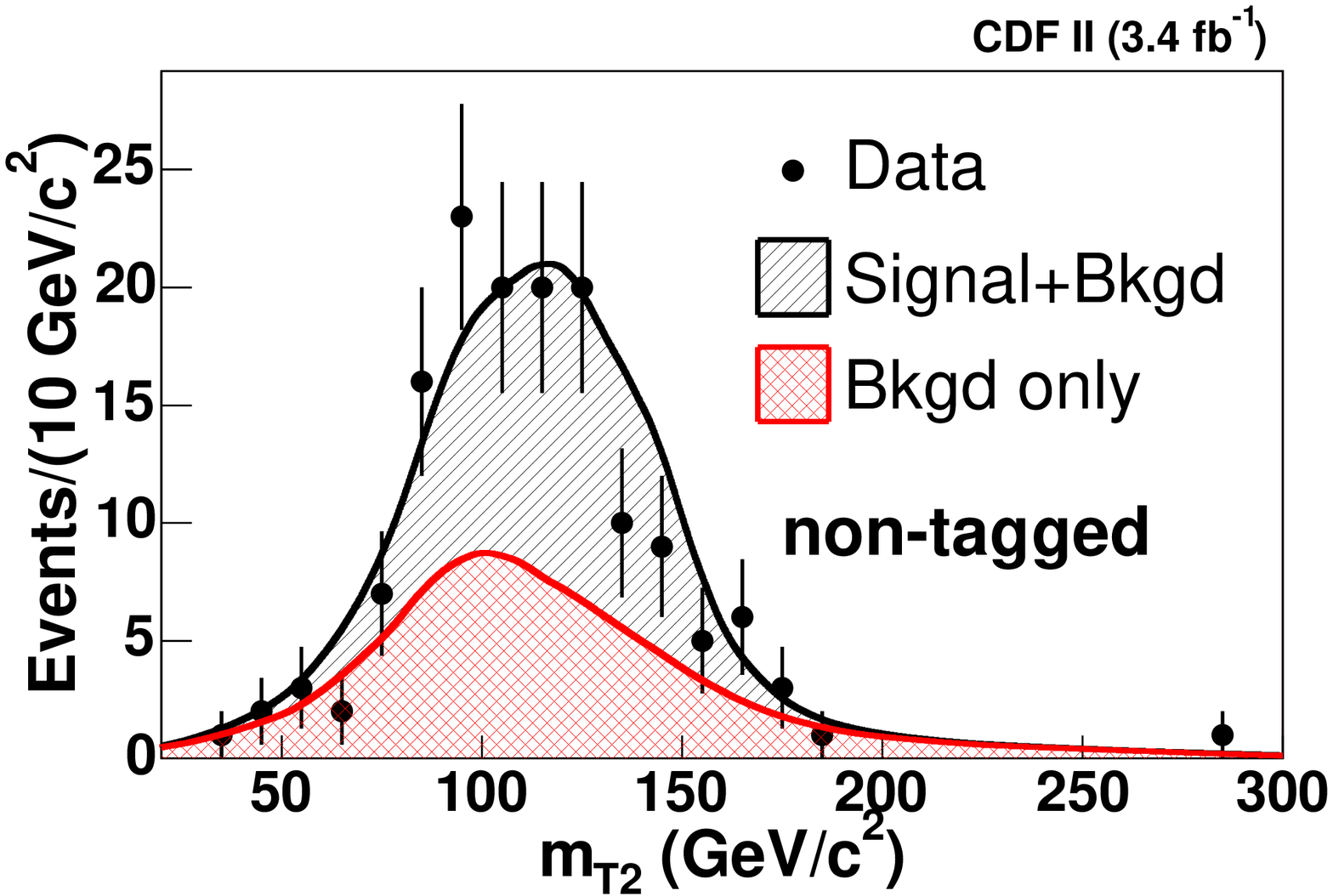}&
\includegraphics[width=0.3\textwidth]{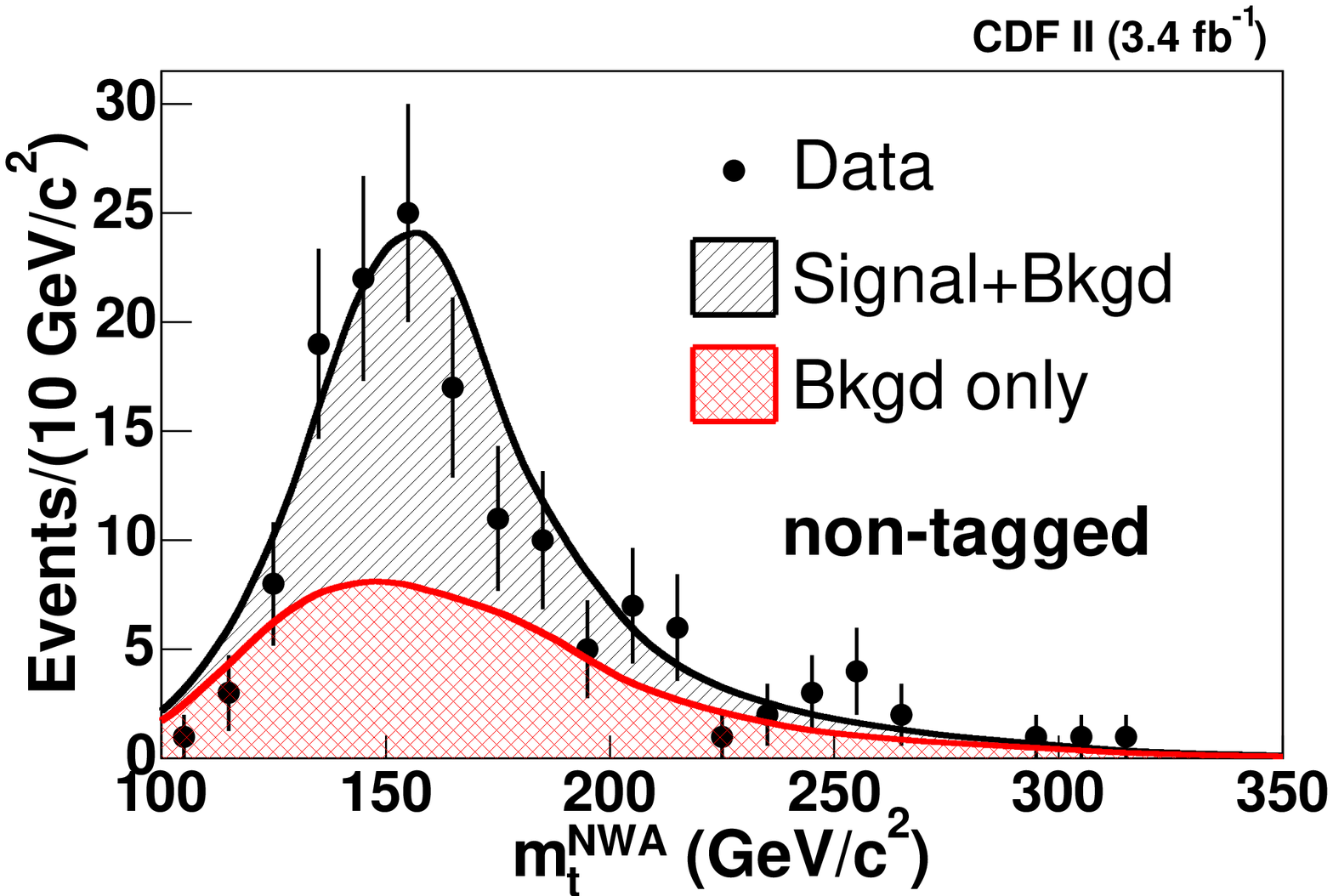}&
\includegraphics[width=0.3\textwidth]{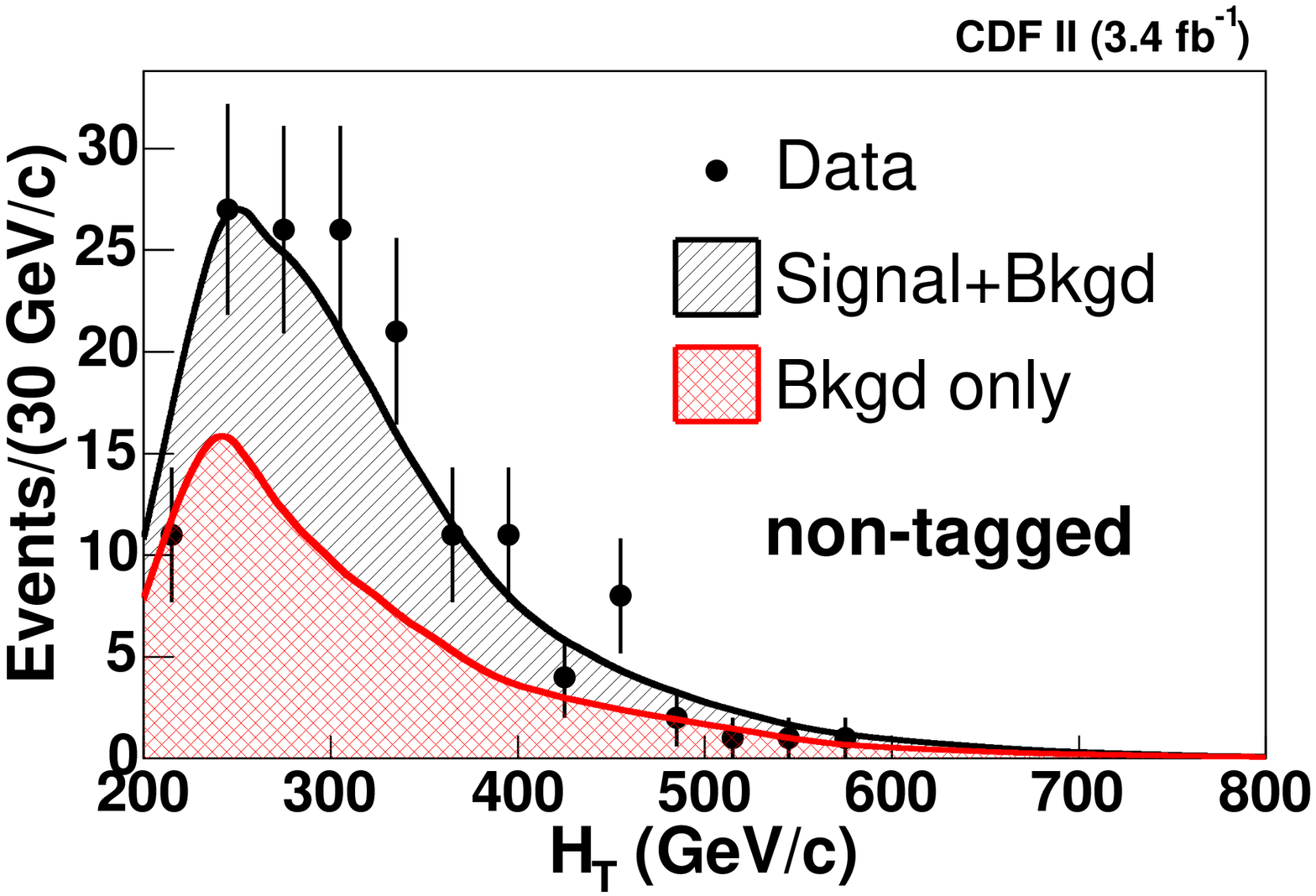}\\
\includegraphics[width=0.3\textwidth]{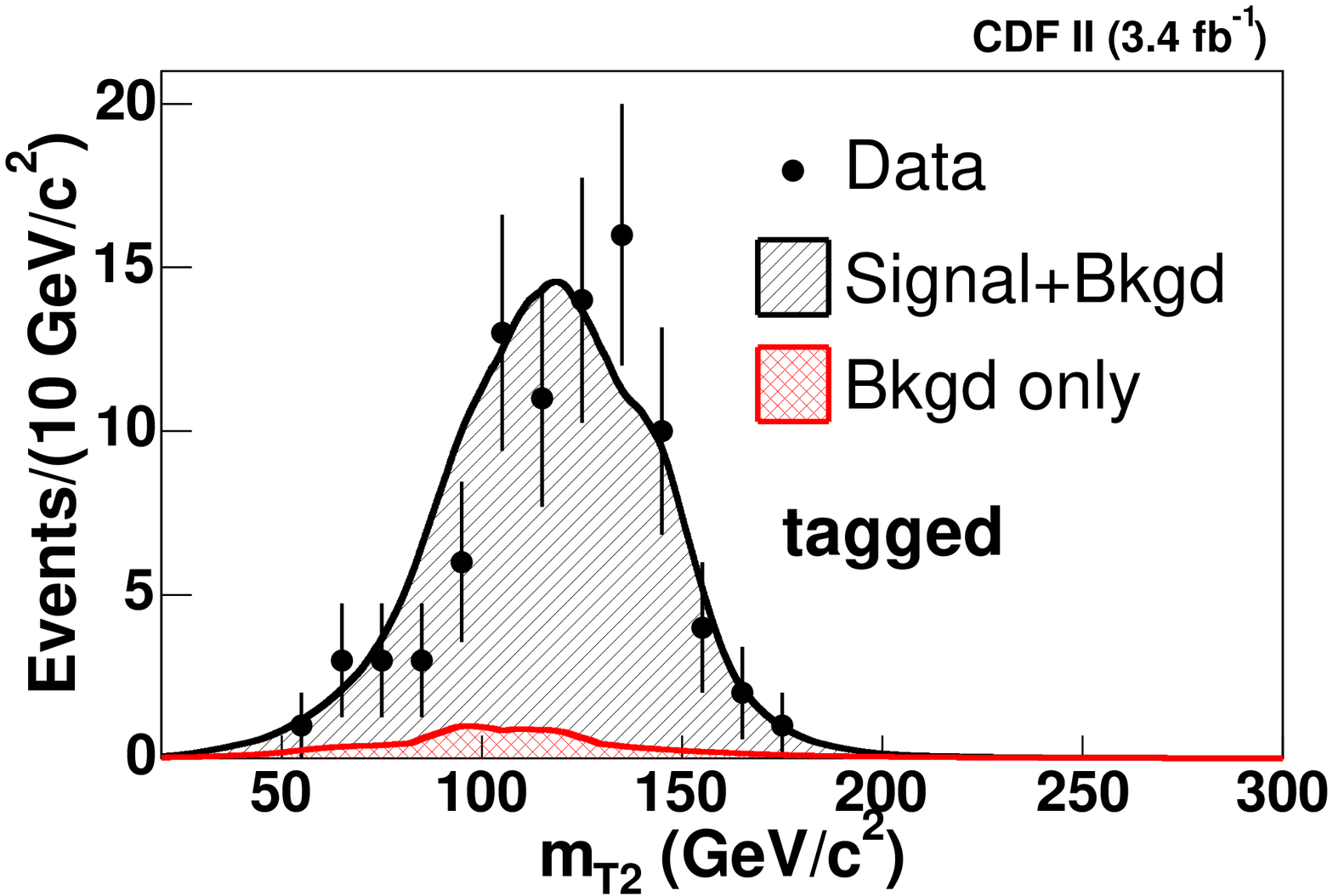}&
\includegraphics[width=0.3\textwidth]{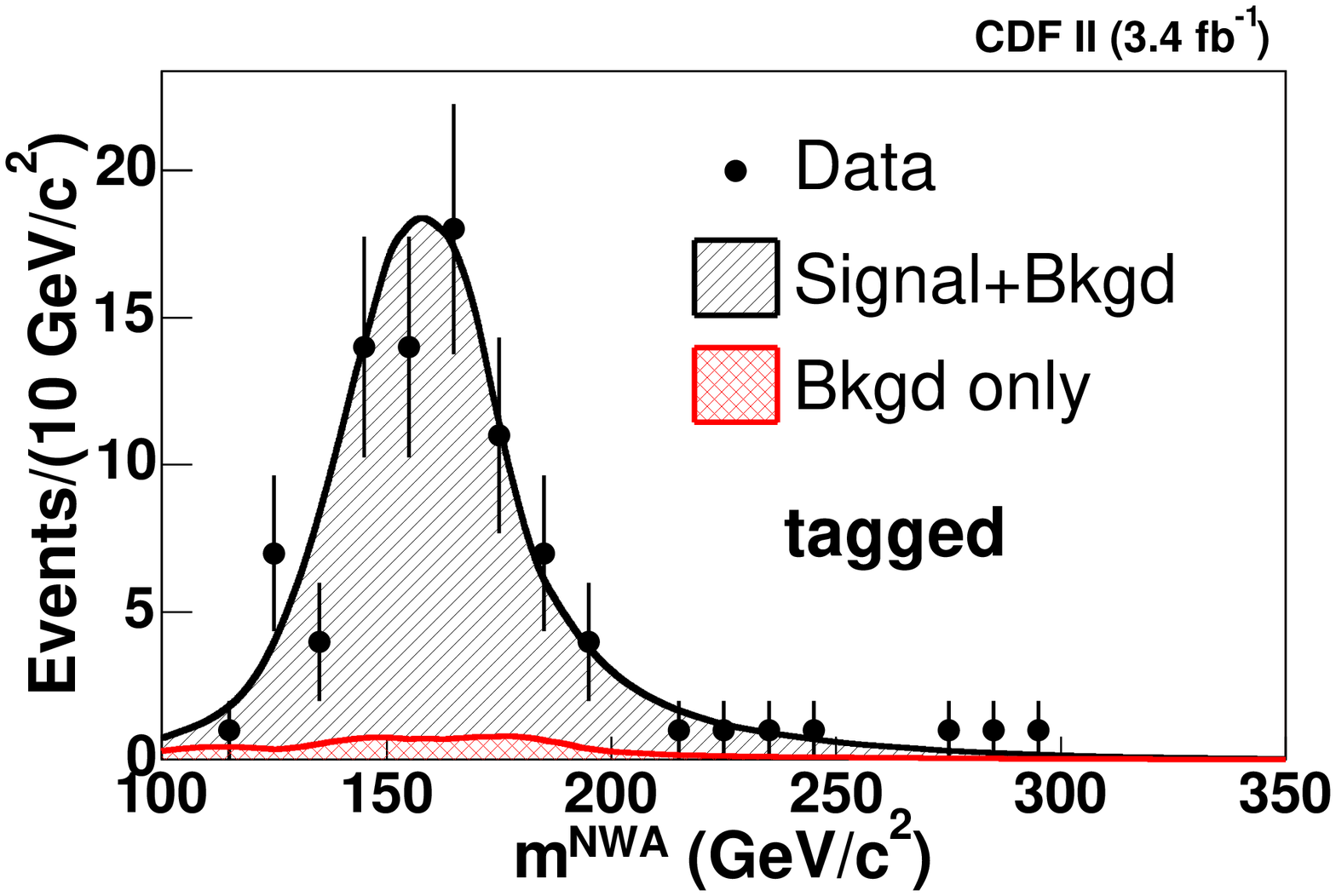}&
\includegraphics[width=0.3\textwidth]{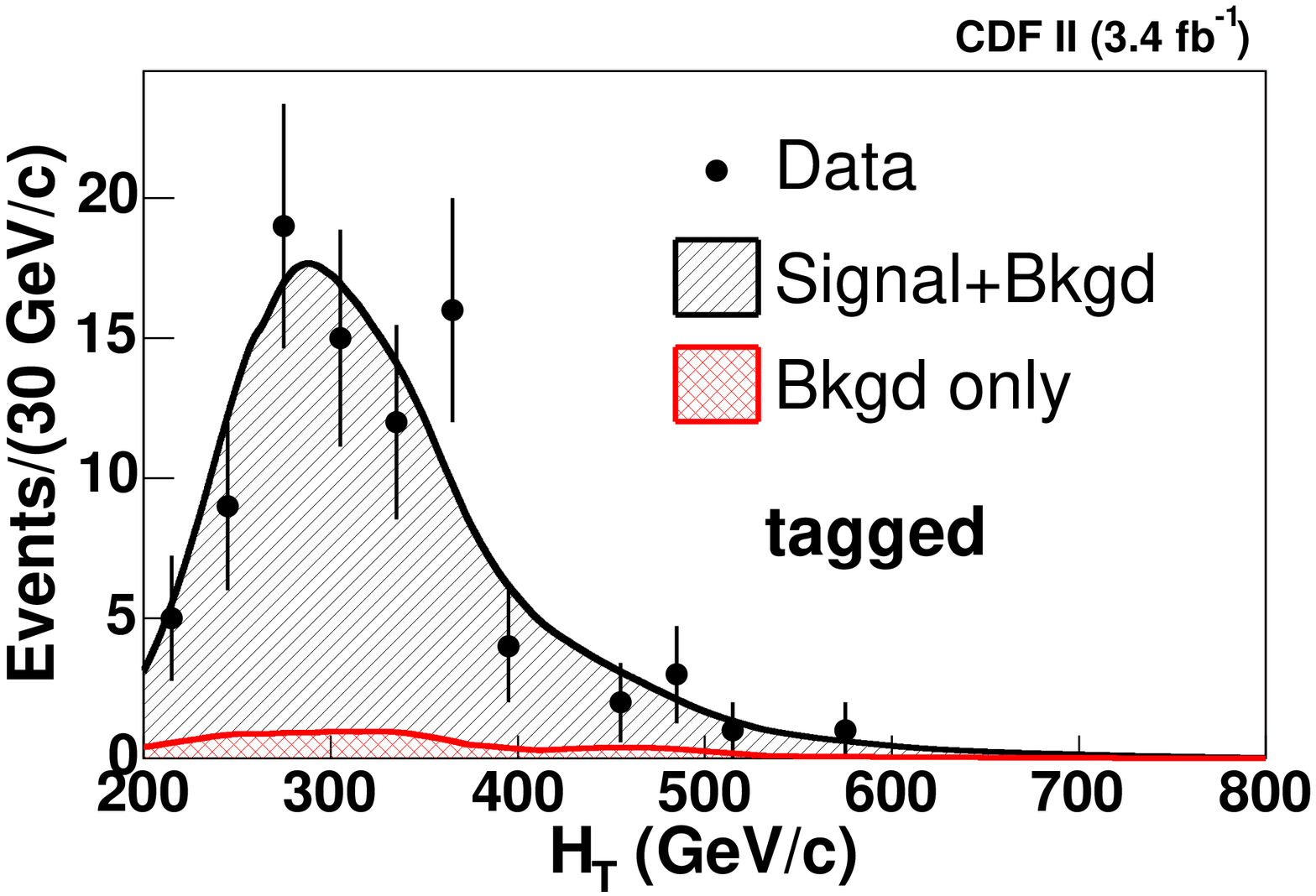}\\
\end{tabular}
\caption[fit result]{Distributions of the three variables used to
estimate the top quark mass, showing the $b$-tagged and non $b$-tagged samples
separately. The data are overlaid with the predictions from the KDE
probability distributions using the top quark mass \mtop =
\gevcc{169} and full background model. 
}
\label{ref:fit}
\end{cfigure1c}

\begin{table*}[ht]
\begin{center}
\caption[Data fit results]{Summary of top quark mass measurements with different
observables. In the right-hand \mtop column, we combine in quadrature the
statistical and systematic uncertainty in order to compare the
precision of the different methods. 
}
\begin{ruledtabular}
\begin{tabular}{ccc}
\label{datafit}
Observables & \mtop (\gevccnoarg) & \mtop (\gevccnoarg)\\ 
\hline
$\ensuremath{{m}_{\text{{T2}}}} $\rule{0pt}{2.6ex} &  \measAStatSyst{168.0}{4.8}{4.0}{2.9} &\measAErr{168.0}{5.6}{5.0}\\
\mtnwa \rule{0pt}{2.6ex}& \measAStatSyst{169.4}{3.3}{3.2}{3.8} &\measAErr{169.4}{5.0}{5.0}\\
$\Ht$ \rule{0pt}{2.6ex}& \measAStatSyst{168.8}{5.1}{6.6}{5.7} &\measAErr{168.8}{7.6}{8.7}\\
\ensuremath{{m}_{t}^{\text{{NWA}}}} \rule{0pt}{2.6ex}and $\ensuremath{{m}_{\text{T2}}} $ 
& \measAStatSyst{169.3}{2.7}{2.7}{3.2} &\measAErr{169.3}{4.2}{4.2}\\
\mtnwa \rule{0pt}{2.6ex}and $\Ht$&  \measAStatSyst{169.6}{2.8}{2.9}{3.8} &\measAErr{169.6}{4.7}{4.8}\\ 
\end{tabular}
\end{ruledtabular}
\end{center}
\end{table*}

In conclusion, we present the top quark mass measurements in the dilepton channel
using $\mTT$. In \invfb{3.4} of CDF data, we measure \mtop using $\mTT$ to be
\begin{linenomath*}
\begin{align*}
\mtop & =  \gevcc{\measAStatSyst{168.0}{4.8}{4.0}{2.9}}\\
      & =  \gevcc{\measAErr{168.0}{5.6}{5.0}},\\
\end{align*}
\end{linenomath*}
and using both \mtnwa and $\mTT$ to be
\begin{linenomath*}
\begin{align*}
\mtop & = \gevcc{\measStatSyst{169.3}{2.7}{3.2}} \\
      & = \gevcc{\measErr{169.3}{4.2}}.\\
\end{align*}
\end{linenomath*}
This is consistent with the most
precise published result in this channel from the CDF~\cite{nndil} and
D0~\cite{d0dil} collaborations. 
We expect further improvements in \mtop with these variables 
as CDF accumulates about a factor of 3 more data during Tevatron Run II.
The measurements in this article are the first application of the $\mTT$ 
variable to data, and demonstrate that $\mTT$ is a powerful 
observable for the mass measurement of the top quark in the dilepton channel. The 
methods described in this article will be applicable to other measurements 
at the Tevatron and soon at CERN's Large Hadron Collider for discriminating 
new physics models and measuring the mass of heavy particles that decay into 
weakly interacting particles such as dark matter candidates.

\begin{acknowledgments}
We thank the Fermilab staff and the technical staffs of the
participating institutions for their vital contributions. This work
was supported by the U.S. Department of Energy and National Science
Foundation; the Italian Istituto Nazionale di Fisica Nucleare; the
Ministry of Education, Culture, Sports, Science and Technology of
Japan; the Natural Sciences and Engineering Research Council of
Canada; the National Science Council of the Republic of China; the
Swiss National Science Foundation; the A.P. Sloan Foundation; the
Bundesministerium f\"ur Bildung und Forschung, Germany; the Korean
Science and Engineering Foundation and the Korean Research Foundation;
the Science and Technology Facilities Council and the Royal Society,
UK; the Institut National de Physique Nucleaire et Physique des
Particules/CNRS; the Russian Foundation for Basic Research; the
Ministerio de Ciencia e Innovaci\'{o}n, and Programa
Consolider-Ingenio 2010, Spain; the Slovak R\&D Agency; and the
Academy of Finland.
\end{acknowledgments}

\bibliography{mT2_paper}

\begin{thebibliography}{29}
\bibitem[{iso(A lepton is isolated if the total $E_{T}$~$(p_{T})$ within a cone
  $\Delta R \equiv \sqrt{(\Delta\eta)^{2}+(\Delta\phi)^{2}}\leq 0.4$, minus the
  lepton $E_{T}$~$(p_{T})$, is $<$10\% of the lepton $E_{T}$~$(p_{T})$ for
  electron~(muon).)}]{isolated}
 \bibinfo{year}{A lepton is isolated if the total $E_{T}$~$(p_{T})$ within a
  cone with $\Delta R \equiv \sqrt{(\Delta\eta)^{2}+(\Delta\phi)^{2}}
= 0.4$ centered on the lepton,
  minus the lepton $E_{T}$~$(p_{T})$, is less than 10\% of the lepton
  $E_{T}$~$(p_{T})$ for electron~(muon).}
\end{thebibliography}
\end{document}